\documentstyle[aps,pre,multicol,epsf]{revtex}

\newcommand{\eps}{\epsilon}
\newcommand{\beq}{\begin{equation}}
\newcommand{\eeq}{\end{equation}}
\newcommand{\bega}{\begin{eqnarray}}
\newcommand{\ega}{\end{eqnarray}}
\newcommand{\vrho}{\varrho}
\newcommand{\non}{\nonumber}
\newcommand{\6}{\partial}
\newcommand{\dx}{\partial_x}
\newcommand{\dt}{\partial_t}
\newcommand{\rmd}{\hbox{d}}
\newcommand{\ra}{\rightarrow}

\newcommand{\eq}[1]{(\ref{eq:#1})}
\newcommand{\Eq}[1]{Eq.~(\ref{eq:#1})}

\newcommand{\tsmq}{\tilde{s}'_{m,q}}
\newcommand{\tsmqp}{\tilde{s}_{m_q,p}}

\newcommand{\smq}{s'_{m_q,q}}
\newcommand{\smqp}{s_{m_{q,p},p}}

\newcommand{\vrmqpp}{\vrho_{m_q}^{'}(p)}
\newcommand{\vrmqp}{\vrho_{m_q}^{'}}
\newcommand{\Tmqpp}{w_{m_q}^{'}(p)}

\begin{document}

\date{\today}
\title{A MultiBaker Map for Thermodynamic Cross-Effects in Dynamical Systems}
\author{L\'aszl\'o M\'aty\'as,$^{(1)}$
Tam\'as T\'el,$^{(1)}$ and
J\"urgen Vollmer$^{(2,3)}$ }
\address{
(1) Institute for Theoretical Physics,
E\"otv\"os University,
P. O. Box 32,
H-1518 Budapest,
Hungary.
\\
(2) Fachbereich Physik, Univ.-GH Essen, 45117 Essen, Germany.
\\
(3) Max-Planck-Institut for Polymer Research,
   Ackermannweg 10, 
   55128 Mainz, Germany.
}
\maketitle
\begin{abstract}
A consistent description of simultaneous heat and particle transport,
including cross effects, and the associated entropy balance is given in the
framework of a deterministic dynamical system. 
This is achieved by a multibaker map where, besides the phase-space density 
{ of the multibaker}, a 
second field with appropriate source terms is included in order to mimic a
spatial temperature distribution and its time evolution. 
Conditions are given to ensure consistency  in an appropriately
defined continuum limit with the thermodynamic entropy balance. 
They leave as the only free parameter of the model 
the entropy flux let directly into a surroundings. 
If it vanishes in the bulk, the transport properties of the model are
described by the thermodynamic transport equations.  
Another choice leads to a uniform temperature distribution. 
It represents transport problems treated by means of a
thermostatting algorithm{ , similar to the one} considered in 
non-equilibrium molecular dynamics.  
\end{abstract}
\draft

\begin{multicols}{2}
\section{Introduction}
Irreversibility in transport models based on dynamical systems with only a
{\em few degrees of freedom\/} have become a subject of intensive recent
studies \cite{Dorf,Focus,Gasp,books,Vance,CELS,ECM,GC,Cohen,Rond,CL}. 
They illustrate how macroscopic transport coefficients are
related to the properties of the microscopic dynamics.
It is a remarkable discovery that in chaotic dynamical systems a
rate of irreversible entropy production can be defined
\cite{books,CELS,ECM,Cohen,Ru,BTV96,ND,Gent,VTB97,VTB98,BTV98,GD,GD2}. 
This development opens the possibility { of} requireing for a consistent 
dynamical-system modelling of an irreversible process the
derivation of both the transport equations and 
the entropy balance. In our approach we shall observe this 
constraint.

Many models \cite{books,Vance,CELS,ECM,GC,Cohen,Rond}
were originally designed to rely on equations of motion, where
transport is induced by an external field and the (average) work
done by the field on the systems is taken out by
a so-called Gaussian thermostat.
This approach { is commonly implemented with periodic boundary 
conditions. It}
has extensively been tested numerically, but 
gives rise to conceptual problems in the interpretation of
the entropy (or heat) flux, 
{ since it only allows to address the global entropy balance 
of a macroscopic system and has no boundaries where entropy fluxes can 
be let into the environment.}
The aim of the present article is to 
investigate a class of
dynamical systems taylored to describe simultaneous { particle}
and heat transport, 
{ driven by appropriate boundary 
conditions and an external field. 
We}
work out {\em local\/} entropy balance, 
{ and identify conditions under which the model can be 
consistent with non-equilibrium thermodynamics.} 

\begin{figure}
    \epsfbox{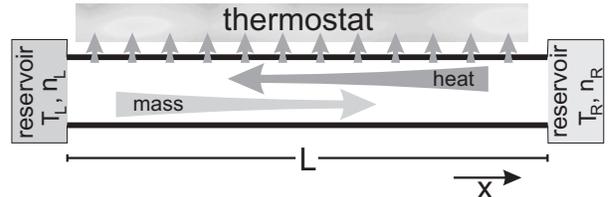} 
\caption[]{\narrowtext
Graphical illustration of the transport process considered.
A system of spatial extension $L$ 
is attached to reservoirs inducing { particle} 
and heat currents 
due to the differences in the densities ($n$) and temperatures ($T$) at the two
ends, as indicated by the arrows. Along the chain heat can be exchanged with a
thermostat. 
\label{fig:geometry}}
\end{figure}

Earlier low dimensional models are devoted exclusively to either 
{ particle (mass,}
charge) transport \cite{Vance,CELS,G,TG,GB2,TVB,KD,Kaufmann}
or to heat conductivity \cite{heat1,heat2}. These transport processes
are driven by a single thermodynamic force. 
It is of basic interest, however, to understand how thermodynamic
cross-effects generated by the presence of two {\em independent}
driving forces can be described in the framework of dynamical
system theory (as for many particle systems see, e.g., \cite{Evans}).
The case of thermoelectric phenomena
in which the driving forces are (i) the temperature difference
and 
(ii) the electric field and/or a density gradient is illustrated by
Fig.~\ref{fig:geometry}. The cross effects imply that 
the temperature gradient contributes to the  electric current, and
the density gradient to the heat current.
Recently, the authors of the present paper suggested an elementary 
model to study these effects \cite{MTV99}. 
Here, we generalize it to explore the conditions under which a
consistency with thermodynamics can be found. 

Part of the above mentioned problems with the entropy balance  
has recently been clarified \cite{VTB97,VTB98,BTV98} in the framework
of multibaker maps modelling quasi one-dimensional { particle}
transport at constant temperature. 
These models are given in terms of the
time evolution of the { (multibaker)} 
phase-space density in a corresponding
two-dimensional {\em single-particle phase space\/}, which in this case
consists of a band of length $L$. 
The fixed width represents a phase-space variable $p$ in
addition to the spatial coordinate $x$ along the { band.} 
The { multi}baker map defines a discrete dynamics (cf.~below), which is 
one-to-one on the band, but does not necessarily preserve the volume locally. 
A recent paper by Tasaki and Gaspard \cite{TG98} shows that analogous
results can be obtained with area-preserving maps by making the width 
of the band position dependent.
This varying height was connected to changes in the { (potential)
energy}, 
and does not appear as a driving
force independent of the density gradient or { the} external field. 

{ Multibaker maps have no natural momentum variable conjugated to the
spatial coordinate $x$. 
Therefore, we characterize the} 
thermodynamic states 
by {\em two independent\/} fields. 
Besides the phase-space density $\varrho$, a new field $w$ is
introduced, whose dynamics describes the evolution of the 
{\em kinetic-energy per particle,\/}, i.e., $\varrho w$ corresponds to 
the kinetic-energy density \cite{consistency}.

To make contact with non-equilibrium thermodynamics, 
the time evolution of average densities in regions of small spatial
extension along the $x$-axis is considered. 
They will be called {\em coarse-grained densities.\/} 
Transport equations in the form of differential equations are obtained 
in a continuum limit, 
where the spatial resolution of coarse
graining is much smaller than the linear size $L$ of the system  and
the time unit $\tau$ of the discrete dynamics is much shorter than that
of the macroscopic relaxations.
In this {\em macroscopic limit\/} the phase-space density $\varrho$ and the
kinetic-energy per particle $w$ are related to the particle density $n$
and to the local temperature $T$, respectively.

It is not obvious that a deterministic dynamical system as simple as a 
multibaker can fulfill all constraints required for consistency with 
thermodynamics \cite{GM} --- not even when taking the macroscopic limit. 
The thermodynamic entropy balance relates the time derivative of the
entropy density $s$ to the entropy production $\sigma^{\rm (irr)}$ per
unit volume and to the entropy flux $\Phi$
\begin{equation}
   \dt s = \sigma^{\rm (irr)} + \Phi.
\label{eq:dts}
\end{equation}
\begin{mathletters}
We consider thermoelectric phenomena 
induced by particles of charge $e$ in a transport process
along the $x$ axis. In a system which is translation invariant
perpendicular to the $x$ axis,
the quantities causing entropy changes can be expressed as \cite{L8} 
\begin{eqnarray}
   \sigma^{\rm (irr)}
&=&
   \frac{e^2 j^2}{\sigma_{el} T}  +  \lambda \left[ \frac{\dx T}{T} \right]^2,
\label{eq:sirr-therm}
\\
   \Phi
& = &
   - \dx j^{(s)} .
\label{eq:Phi-therm}
\end{eqnarray} 
\label{eq:sirr-Phi-therm}
\end{mathletters}
Here $\sigma_{el}$ and $\lambda$ denote the electric and the heat
conductivity, respectively, and 
\begin{mathletters}
\begin{eqnarray}
j & = & n v_{el} -\frac{\sigma_{el}}{e^2}(\dx \mu_c + e \alpha \dx T) ,
\label{eq:j-macr}
\\
   j^{(s)} & = & \frac{e \Pi}{T} j - \lambda \frac{\dx T}{T}
\label{eq:js-macr}
\end{eqnarray}
\label{eq:therm-currents}
\end{mathletters}
are the particle and the entropy current densities, respectively.
In \Eq{therm-currents}, $\mu_c$ denotes the chemical potential of
the particles, and $v_{el}$ is the drift velocity due to an
external electric field $E$, which is related to $\sigma_{el}$ by 
\begin{equation}
   v_{el} = \frac{\sigma_{el} E}{e n} .
\label{eq:vs}
\end{equation}
$\Pi$ is the Peltier coefficient, and $\alpha$ the thermoelectric power 
(or Seebeck coefficient). 
Thermodynamic cross effects are manifest in (\ref{eq:therm-currents}), 
and the corresponding Onsager
relations imply a relation between the transport coefficients 
$\alpha$ and $\Pi$.

Since the entropy plays a central role in these relations, one has to 
choose an appropriate entropy concept for the multibaker. 
One obvious candidate is the Gibbs entropy $S^{(G)}$ defined with
respect to the phase-space density $\varrho$.  
Due to the ever refining phase-space structures, 
the { chaotic dynamics generates from every} 
smooth initial density, this entropy {\em  never\/}  
becomes time-independent, not even in a macroscopically 
steady state.  In contrast, a corresponding entropy $S$, whose definition is
based on coarse-grained densities, is thermodynamically well-behaved and its
value per unit length is the analog of the entropy density $s$ appearing in
(\ref{eq:dts}). 
This entropy will be called the {\em coarse-grained
  entropy\/}.  The irreversible entropy production of arbitrary steady and
non-steady states was identified as the time derivative of the difference
between the Gibbs and the coarse-grained entropy \cite{VTB98,BTV98}.

Although the thermodynamic relation (\ref{eq:Phi-therm}) requires the entropy
flux to be the divergence of the entropy current, we allow for deviations from
thermodynamics in that  
we do not exclude the presence of an additional term.  
This term is interpreted as the consequence of a thermostat, which can
{\em locally\/} remove or release heat, leading to an additional entropy flux
$\Phi^{\rm (thermostat)}$.  
In such cases 
\begin{equation} 
   \Phi = - \dx j^{(s)} + \Phi^{\rm (thermostat)}.
\label{eq:Phi-thermo} 
\end{equation}
We call a system thermostatted, whenever $\Phi^{\rm (thermostat)}$ differs 
from zero. 
Depending on the details of the model, we are able to study both
non-thermostatted and thermostatted 
systems, and in the latter case we shall be able  
{ generate arbitrary stationary temperature profiles.} 
To our knowledge, these features have not yet been explored in
non-equilibrium molecular dynamics simulations \cite{books,Evans}. 

The paper is organized as follows.
In Sect.~\ref{sec:Thermodynamics} we revisit the thermodynamics of
irreversible processes by rewriting the expressions of entropy
production as well as of particle and entropy currents in forms amenable
to a comparison with the results of multibaker maps.
Subsequently, (Sect.~\ref{sec:MBaker}) we introduce the multibaker
map, and discuss the time evolution
of the phase-space density and the kinetic-energy density.
The Gibbs and coarse-grained entropies and their dynamics 
are studied in Sect.~\ref{sec:entropies}. 
Subsequently, in Sect.~\ref{sec:limit} the macroscopic limit of the obtained
expressions is taken.  
Conditions on the baker dynamics to make it consistent with thermodynamics are
explored in Sect.~\ref{sec:compare}.  
A short discussion is devoted to thermostatted cases (Sect.~\ref{sec:therm}),
and our main results are summarized in Sect.~\ref{sec:conclude}. 
The paper is augmented by two appendices.  
App.~\ref{app:definition} is devoted to a formal definition of the map and the
resulting time evolution of the densities.   
In App.~\ref{app:stability} it is shown that the macroscopic results do not
depend on the prescription for coarse graining. 

\section{Non-Equilibrium Thermodynamics}
\label{sec:Thermodynamics}
In this section we recall the thermodynamic description of  
transport induced by two independent driving fields \cite{GM}. 
The most general situation is treated, which comprises the 
presence of an external field, as well as gradients in the particle
density and the temperature.

\subsection{Thermodynamic forces and currents}
We consider a system of particles of charge $e$ in a nonmoving
background.
Due to an (electro-) chemical potential gradient and a
temperature gradient, both a particle and an energy current is
flowing through the system.
Let $j$ and $j^{(u)}$, respectively, denote the density of
these currents in a  frame of reference fixed to the background.
In this setting the number density of particles $n$ and the density
$u$ of the total energy are locally preserved.
(The density $u$ also contains the potential energy due to an
external field and the kinetic energy
of the ordered motion.)

In order to derive the entropy balance in a region of fixed volume,
we start from the 
conservation laws
\begin{mathletters}
\begin{eqnarray}
   \dt n &=& - \nabla j,
\label{eq:cn}
\\
   \dt u &=& - \nabla j^{(u)},
\label{eq:cu}
\end{eqnarray}
and express 
the time derivative of the entropy density per unit volume $s$
\begin{equation}
   \dt s = - \nabla j^{(s)} + \sigma^{\rm (irr)} 
\label{eq:cs}
\end{equation}
\end{mathletters}
in terms of the entropy-current density $j^{(s)}$ and the irreversible
entropy production $\sigma^{\rm (irr)}$. 

Considering a system in local equilibrium, the Gibbs relation
\begin{equation}
   T \,\rmd s = \rmd u - \mu \,\rmd n,
\label{eq:fl}
\end{equation}
holds with $T$ and $\mu$ denoting the local 
temperature and electro-chemical potential, respectively.
To find the time derivative of $s$, 
we write the local temporal change of (\ref{eq:fl})
\begin{equation}
   \dt s
=     
   \frac{1}{T}  \dt u
  - \frac{\mu}{T} \dt n
\label{eq:st}
\end{equation}
in the form of (\ref{eq:cs}) by identifying 
\begin{mathletters} 
\begin{equation}
   j^{(s)}=\frac{j^{(u)}- \mu j}{T}
\label{eq:ju-def}
\end{equation}
as the total entropy current, and
\begin{equation}
   \sigma^{\rm (irr)}
=
   j^{(u)} \nabla \frac{1}{T} - j \nabla \frac{\mu}{T}
=
   -(j^{(s)} \nabla {T} + j \nabla \mu) \frac{1}{T}
\label{eq:sig}
\end{equation}
as the irreversible entropy production.
\end{mathletters} 

Equation (\ref{eq:sig}) shows that the currents $j$ and $j^{(s)}$ 
are conjugate to the thermodynamic forces  $-\nabla \mu/T$ and
$-\nabla {T}/T$,
respectively. Therefore, in the spirit of non-equilibrium thermodynamics,
these currents can be expressed in terms of the forces as
\begin{mathletters}
\begin{eqnarray}
   j  &=& - L_{11} \frac{\nabla\mu}{T} - L_{12} \frac{\nabla T}{T},
\label{eq:j-th1}
\\[2mm] 
   j^{(s)} &=& - L_{21}  \frac{\nabla\mu}{T} - L_{22} \frac{\nabla T}{T}
\label{eq:js-th1}
\\[1mm] 
&=& - \frac{L_{11} L_{22} - L_{12} L_{21}}{L_{11}}
            \frac{\nabla T}{T}
      +  \frac{L_{21}}{L_{11}}   j  , 
\label{eq:jsthermo}
\end{eqnarray}
\end{mathletters}
where the latter expression for $j^{(s)}$ was obtained by inserting 
\eq{j-th1} into \eq{js-th1}. 
Moreover, by inserting (\ref{eq:js-th1}) into (\ref{eq:sig}), one
expresses the irreversible entropy production as
\begin{eqnarray}
   \sigma^{\rm (irr)}
& = &
   \frac{L_{11} L_{22} - L_{12} L_{21}}{L_{11}}
            \left( \frac{\nabla T}{T} \right)^2
      +  \frac{j^2}{L_{11}} 
\non \\
&&
      +  \frac{L_{12}-L_{21}}{L_{11}} \frac{\nabla T}{T} j .
\label{eq:production}
\end{eqnarray}
After using the Onsager relation $L_{12} = L_{21}$,
this expression for the irreversible entropy production is a
quadratic form, which takes only nonnegative values provided that
the matrix of kinetic coefficients is positive definite,
i.e., the well-known condition
   $L_{11} L_{22} - L_{12} L_{21} > 0$
is fulfilled \cite{GM,Prigogine}.
 
\subsection{Identifying transport coefficients}
It is worth expressing the kinetic coeffcients $L_{ij}$
by means of directly measurable quantities.
The total electro-chemical potential can be split
as $\mu = \mu_c + e \phi$, where $\mu_c$ is the chemical part,
and $\phi$ is the electric potential.
Since $E = -\nabla\phi$ and $j_{el}=ej$ is the electric current, 
we find that $L_{11}$ is
proportional to the electric conductivity $\sigma_{el}>0$:
\begin{mathletters}
\begin{equation}
   L_{11} = \frac{\sigma_{el} T}{e^2}.
\end{equation}

In the absence of a particle current (i.e., for $j = 0$),
$T j^{(s)}$ provides the heat current, so that in view of (\ref{eq:jsthermo}) 
\begin{equation}
   \frac{L_{11} L_{22} - L_{12} L_{21}}{L_{11}}  = {\lambda},
\end{equation}
where $\lambda > 0$ is the heat conductivity.

At zero particle current and constant chemical potential,  
a temperature gradient induces an electric field, which is  
conventionally written as $\alpha \nabla T$, where $\alpha$ is called 
the thermoelectric power (or the Seebeck coefficient). 
Consequently, from (\ref{eq:j-th1}) one finds 
\begin{equation}
   L_{12} = \frac{\alpha \sigma_{el}}{Te}.
\end{equation}

Finally, in a sytem without temperature gradients,
the entropy current due to the presence of an electric current $ej$
amounts to $\Pi e j/T$, where $\Pi$ is the Peltier coefficient.
Hence, Eq.~(\ref{eq:jsthermo}) implies 
\begin{equation}
   L_{21} = \frac{\Pi \sigma_{el}}{e}.
\end{equation}
\label{eq:ells}
\end{mathletters}

Using the phenomenological coefficients (\ref{eq:ells}) we write 
\begin{mathletters}
\begin{eqnarray}
   j &=& -\frac{\sigma_{el}}{e^2}(\nabla \mu + e \alpha \nabla T) ,
\label{eq:j-th}
\\
   j^{(s)} &=&  - \lambda \frac{\nabla T}{T}
            +  \frac{e\Pi}{T} j.
\label{eq:js-th}
\end{eqnarray}
\end{mathletters}
Note that the Onsager relation $L_{12}=L_{21}$ makes the Peltier
and Seebeck coefficients connected as
\begin{equation}
   \Pi = T \alpha.
\label{eq:On}
\end{equation}
Substituting Eqs.~(\ref{eq:ells}) into (\ref{eq:production}) one thus derives
(\ref{eq:sirr-therm}).

\subsection{Relating transport and diffusion coefficients}
It is worth replacing the chemical potential in the
expressions for the currents and entropy production by the density $n$
and temperature $T$. 
We write
\begin{equation}
   \nabla \mu_c= \frac{D e^2}{\sigma_{el}} \nabla n - \varsigma \nabla T,
\end{equation}
where the diffusion coefficient is defined as
\begin{mathletters} 
\begin{equation}
   D
=
   \frac{\sigma_{el}}{e^2}
\left.
   \left( \frac{\partial \mu_c}{\partial n}
   \right)
\right|_T  ,
\label{eq:Diffusion}
\end{equation}
 and
\begin{equation}
   \varsigma
\equiv
-  \left.
      \left( \frac{\partial \mu_c}{\partial T}  \right)
   \right|_n
\label{eq:varsigma} 
\end{equation}
\end{mathletters} 
is a quantity of dimension entropy per particle. 
By means of the drift velocity $v_{el}$ [\Eq{vs}] one can rewrite the 
particle current density (\ref{eq:j-th}) as 
\begin{equation} \label{eq:j1}
   j=v_{el} n  - D \nabla n - k\frac{nD}{T} \nabla T
\end{equation}
where  
\begin{equation}
k\frac{nD}{T} \equiv \frac{\sigma_{el}}{e^2} \; (e \alpha - \varsigma) .
\label{eq:k}
\end{equation}
The entropy current then takes the form
\begin{equation}     \label{eq:jsthermo1}
   j^{(s)}=
-\left(\lambda + knD \frac{e\Pi}{T}  \right) 
\frac{\nabla T}{T}+
   \frac{ e \Pi}{T}( v_{el} n  -  D \nabla n) ,
   \label{eq:ecur}
\end{equation}
and in view of \eq{On}, \eq{Diffusion} and \eq{k} one obtains 
\begin{equation}
e \Pi = k n \left.
   \left( \frac{\partial \mu_c}{\partial n}
   \right)
\right|_T  - T 
\left.
      \left( \frac{\partial \mu_c}{\partial T}  \right)
   \right|_n .
\label{eq:Pe}
\end{equation}
Note that in the entropy current (\ref{eq:ecur}) 
the coefficients in front of the $v_{el} n$ and the 
$-D\nabla n$ terms coincide. 

We finally 
mention that by taking the limit $E, e \rightarrow 0$ at constant 
$e \Pi, e \alpha$, and $\sigma_{el}/e^2$, the thermoelectric
problem  is formally mapped onto the problem of thermal diffusion in a binary
mixture. 
In that case $n$ stands for the concentration of one of the diffusing
materials, and the quantity $kD$ is the thermal diffusion coefficient
\cite{L6}. 
Based on this analogy, we consider $kD$ in \eq{k} as the thermal diffusion
coefficient of charged particles in the thermoelectric problem.

\begin{figure}
    \epsfbox{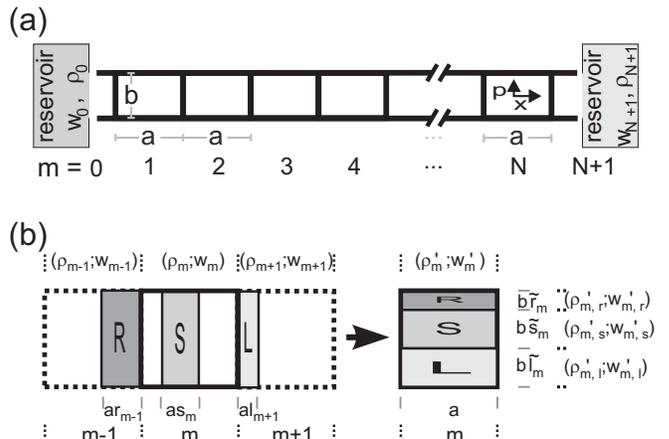} 
\caption[]{\narrowtext
Graphical illustration of the action of the multibaker map 
on the phase space $(x,p)$. 
(a) The mapping is defined on a domain of $N \equiv L/a$ identical
rectangular cells of size $a \times b$, with boundary condition
imposed in two additional cells $0$ and $N+1$.
(b) 
After every time unit $\tau$ each cell is divided into three columns, 
which are squeezed and stretched to obtain
horizontal strips of width $a$.
The average values of the fields 
$\varrho(x,p)$ and $w(x,p)$ on the cells (strips) 
[cf. Eqs. (\ref{eq:nm'}) and (\ref{eq:wm'})] are given 
on the margins. 
Iteration of this rule defines the time evolution. 
\label{fig-mbaker}}
\end{figure}

\section{The Multibaker Map}
\label{sec:MBaker}
Baker maps are known to be prototypes of strongly chaotic systems
\cite{Ott93}.
Multibaker maps are a generalization, where a spatially extended
system is modelled by a chain of mutually interrelated baker maps,
in order to model 
transport \cite{G,TG,TVB,VTB97,VTB98,BTV98,TG98}
via the dynamics of the { (multibaker)} phase-space density $\varrho$. 
The single-particle phase space 
modelled by the multibaker map consists of
$N$ identical cells of width $a$  and height $b$, which are
labelled by the index $m$ (Fig.~\ref{fig-mbaker}a).
After each time unit $\tau$ every cell is divided into three columns
(Fig.~\ref{fig-mbaker}b). 
The right (left) column of width
     $a r_{m}$
   ( $a l_{m}$ )
is mapped onto a strip of width $a$ and of height
     $b \widetilde{r_{m+1}}$
   ( $b \widetilde{l_{m-1}}$ )
in the right (left) neighbouring cell.
The middle one preserves its area
   $\widetilde{s_m} = s_m$.
The map is {\em globally\/} phase-space preserving, i.e.,
   $s_m+r_m+l_m = \widetilde{r_m}+\widetilde{l_m}+\widetilde{s_m} = 1$.
A formal definition is given in App.~\ref{app:definition}.
The map mimics the time evolution of a many-particle 
system with weak interactions 
in a single particle phase-space 
\cite{weak}. 
The equilibrium equations of state will turn out to be those of a
classical ideal gas. 

The $(x,p)$ dynamics of the multibaker map is considered to be a 
{\em microscopic \/} dynamics in the sense that it is deterministic, 
chaotic, and mixing. 
It drives two fields which generate ever refining phase-space structures. 

Besides these fields, 
we also consider coarse-grained densities obtained by averaging over 
the cells.  
The cell width $a$ is considered to be the smallest length scale where a
thermodynamic description applies.  
In the spirit of local thermodynamic equilibrium, the local averages of the
microscopic variables characterize the  
thermodynamic state in the cells. 
The temporal evolution of the coarse-grained versions of $\varrho$ and
$w$ is consequently expected to describe an approach towards a steady
state, { where the} 
the coarse-grained fields do { no longer} change in time 
{ (in contrast to the fully resolved fields, which 
approach closer and closer towards fractal distributions).}
To emphasize that coarse-graining is taken over the cells, the
coarse-grained density will also be called the {\em cell  density\/}. 

The dynamics of earlier 
multibaker models is the same for all cells.  There
can be inhomogeneities in the densities, but the evolution equations are kept
translation invariant. 
Here, we relax this constraint by { allowing the transfer rates of 
cell $m$ to depend on the coarse grained 
fields in cell $m$ and its neighbors.}  
This dependence mimics the effect of the thermodynamic driving force due to
{ for instance} a local temperature gradient.  
It induces an $m$-dependence of the parameters. 
Since all calculations can be performed without ever referring to the form of
these dependencies, we will not yet specify them but start from a
map with a general set of space and time-dependent parameters.  
Their form will be fixed a posteriori by comparison with thermodynamics. 

Due to the self similarity of the multibaker dynamics, the local
transport and entropy balance can be worked out by a calculation considering
one time step only. 
A discussion of more general prescriptions for the coarse graining is
relegated to  App.~\ref{app:stability}. 

\subsection{Evolution of the phase-space density}

Thermodynamic transport equations describe the time evolution of  
the phase-space density $\varrho$ and the kinetic-energy density $w$. 
For explicit calculations of their time evolution we always start with the
constant values $\varrho_m$ and $w_m$ in every cell $m$. 
This is convenient from a technical point of view, and does not lead to a
principal restriction of the domain of validity of the model as it will be
demonstrated in App.~\ref{app:stability}. 
After one step of iteration the densities will be piecewise constant on the
strips $i=R,S,L$ defined in Fig.~\ref{fig-mbaker}b. 
Due to the conservation of the particle number, they are 
\begin{eqnarray}
   \varrho_{m,r}' & = & \frac{r_{m-1}}{\widetilde{r_{m}}} \; \varrho_{m-1} ,
\non \\
   \varrho_{m,s}' & = & \varrho_{m} ,
\label{eq:nm'}
\\
   \varrho_{m,l}' & = & \frac{l_{m+1}}{\widetilde{l_{m}}} \; \varrho_{m+1} .
\non
\end{eqnarray}
The factors
   ${r_{m-1}}/{\widetilde{r_{m}}}$
and
   ${l_{m+1}}/{\widetilde{l_{m}}}$
give rise to local contraction or expansion of phase-space 
volumes.  

The coarse-grained density $\varrho'_m$ after one time step
is the average of the contributions \eq{nm'} on the different strips:
\begin{equation}
   {\varrho'_m}
=
   (1-r_m-l_m) \varrho_m + r_{m-1} \varrho_{m-1} + l_{m+1} \varrho_{m+1} .
\label{eq:rhoprime}
\end{equation}
Multiplying the equation by $\tau^{-1}$ and introducing
the current 
\begin{eqnarray}
   j_{m}
=
   \frac{ab}{\tau}(r_{m}\varrho_{m}-l_{m+1}\varrho_{m+1}) 
\label{eq:jm+}
\end{eqnarray}
through the right boundary of cell $m$, 
Eq.~(\ref{eq:rhoprime}) appears in the form of a 
continuity equation 
\begin{equation}
   \frac{b(\varrho'_m-\varrho_m)}{\tau}
=
   -\frac{j_m-j_{m-1}}{a} .
\label{eq:dt_varrho'}
\end{equation}
It can be seen as the discrete counterpart of (\ref{eq:cn}).
Note that by definition the current 
through the left boundary of cell
$m$ is the same as the current  
through the right boundary of cell $m-1$. 
Other types of currents associated with cell $m$ will also be defined as
a flow through the right boundary.   

\subsection{Evolution of the kinetic-energy density}
The $(x,p)$ dynamics does not imply any constraint on $w$. 
Its time evolution can be chosen according to physical intuition. 
In contrast to the particle density, we consider the kinetic energy per unit
volume $\varrho_m w_m$ as a non-conserved quantity.  
Besides a contribution from the particle flow, the new values $w'_{m,i}$ on
the strips $i=R,S,L$ will therefore contain terms characterized by 
a local source strength $q_m$, which accounts for a local heating: 
\begin{eqnarray}  
   \varrho_{m,r}' \, w_{m,r}'
& = &
   \frac{r_{m-1}}{\widetilde{r_{m}}} \, \varrho_{m-1} \, w_{m-1} \;
   \left[ 1 + \tau q_m \right] ,
\non \\
   \varrho_{m,s}' w_{m,s}'
& = &
   \varrho_m \, w_{m}
   \left[ 1 + \tau q_m \right] ,
\label{eq:wm'}
\\
   \varrho_{m,l}' \, w_{m,l}'
& = &
   \frac{l_{m+1}}{\widetilde{l_{m}}} \, \varrho_{m+1} \, w_{m+1}
   \left[ 1 + \tau q_m \right] .
\non
\end{eqnarray}
The source term $q_m$ is taken constant in every cell since 
more general coices only effect terms which
drop out in the macroscopic limit. 
The particular form of $q_m$ will only be specified later, in order to
demonstrate that there is a {\em unique\/} choice for $q_m$, where the time
evolution of the kinetic energy can become 
consistent with thermodynamics. 
Other choices for the source term $q_m$ lead to a non-vanishing entropy flux
$\Phi^{\rm (thermostat)}$, and can be considered to characterize a
thermostat in the spirit of non-equilibrium molecular dynamics. 

An update of the kinetic-energy density can be calculated similarly to the
update of the particle density.  
In cell $m$ the average value $\varrho'_m w'_m$ 
after one time step is obtained by averaging the different 
contributions on the strips in cell $m$ [cf.~Eq.~(\ref{eq:wm'})], 
yielding 
\begin{eqnarray}
   \varrho_m^{'} w_m^{'}
&=&
   [\varrho_m w_m(1-r_m-l_m) + r_{m-1}\varrho_{m-1}w_{m-1}
\non\\
&&
   + l_{m+1}\varrho_{m+1}w_{m+1}] \; [1+\tau q_m].
\label{eq:wmprime1}
\end{eqnarray}
For a fixed set of transition probabilities $r_m, s_m, l_m$ and
$q_m = 0$, Eq.~(\ref{eq:wmprime1}) amounts to a passive advection of the field
$w$ by the deterministic dynamics.  
The possible dependence of the transition probabilities on differences of the
coarse-grained $w$ and the presence of the source $q_m$, however, make the
advection nonpassive. 

Equation~(\ref{eq:wmprime1}) can also be rewritten in the form of the discrete
balance equation 
\begin{eqnarray}
   \frac{b(\varrho_m^{'} w_m^{'} - \varrho_m w_m)}{\tau}
=
   b \varrho_m^{'} w_m^{'} \frac{q_m}{1 + \tau q_m}
 -  \frac{j_m^{(\varrho w)} - j_{m-1}^{(\varrho w)}}{a} .
\label{eq:dt_varrho_m-w} 
\end{eqnarray}
The first term on the right-hand side characterizes the source strength of the
field $\varrho_m w_m$ per unit time $\tau$, and the second one is the discrete
divergence of the $\varrho w$-current 
\begin{equation}
   j_m^{(\varrho w)}
=
   \frac{ab}{\tau} (r_{m}\varrho_{m} w_{m} - l_{m+1}\varrho_{m+1} w_{m+1}) .
\end{equation}
Since $\varrho w$ plays the role of a kinetic-energy, we consider 
$j^{(\varrho w)}$ to be an energy current.

\subsection{Diffusion and drift}
The local transition probabilities
   $r_m$, $s_m$ and $l_m$
govern the evolution of the coarse-grained densities $\varrho_m$ and $w_m$. 
In view of the master equations \eq{rhoprime}, the cell-to-cell dynamics of
the model is equivalent to the dynamics of 
random walkers with fixed step lenght $a$ and local transition probabilities $r_m$ and $l_m$ over time unit $\tau$. 
Such random walks are characterized \cite{Reif} by the local drift  $v_m$ and
diffusion coefficient $D$: 
\begin{mathletters}
\begin{eqnarray}
   r_{m} - l_{m} & =& \frac{\tau}{a} v_m  ,
\\
   r_{m} + l_{m} &=& \frac{2\tau}{a^{2}}D .
\end{eqnarray}
\label{eq:rm+-lm}
\end{mathletters}
Hence, the transition probabilities $r_m$ and $l_m$ can be expressed as
\begin{mathletters}
\begin{eqnarray}
   r_m &=& \frac{\tau D}{a^2} \left(1+\frac{av_m}{2D} \right) ,
\\
   l_m &=& \frac{\tau D}{a^2} \left( 1-\frac{av_m}{2D} \right) .
\end{eqnarray}
\label{eq:rmlm}
\end{mathletters}
We allow in the present paper only for a location dependence of the drift
$v_m$, but keep the diffusion coefficient spatially homogenous. 
The $m$-dependence of the drift should be a consequence of the inhomogeneity
of the kinetic-energy (temperature) gradient along the chain. 

In spite of the freedom we still have in specifying $v_m$, 
these definitions already allow us to write the currents in 
a form very close to their thermodynamic counterparts.
The current for the phase-space density appears in the form
\begin{mathletters}
\begin{equation}
   j_m = \frac{b}{2} (\varrho_m v_m + \varrho_{m+1} v_{m+1})
      - b D \; \frac{\varrho_{m+1} - \varrho_m}{a}
\label{eq:jm+b} 
\end{equation}
a discrete version of \eq{j-macr}.
Similarly, for the $\varrho w$-current one obtains
\begin{equation}
   j_m^{(\varrho w)}
=
   w_m j_m -  b\,\varrho_{m+1} D \frac{w_{m+1}-w_m}{a} \;
      \left[ 1 - \frac{a v_{m+1}}{2D} \right]
\label{eq:jmw+b} 
\end{equation}
\label{eq:j+b} 
\end{mathletters}
which describes an advection of the $w$ fields by the particle current 
and a contribution from the discrete gradient of the kinetic energy.

\subsection{Parametrizing phase-space contraction}
\label{sec:parametrize} 
Due to the condition
   $\widetilde{r_m} + \widetilde{l_m} = r_m + l_m$, 
one can express
   $\widetilde{r_{m}}$ and $\widetilde{l_{m}}$
in an analogous way as (\ref{eq:rmlm}) by introducing an additional 
parameter $\eps$ via 
\begin{mathletters}
\begin{eqnarray}
   \widetilde{r_m}
&=&
   \frac{\tau D}{a^2} \left( 1+\eps\frac{av_m}{2D} \right),
\\
   \widetilde{l_m}
&=&
   \frac{\tau D}{a^2} \left( 1-\eps\frac{av_m}{2D} \right) .
\end{eqnarray}
\label{eq:trmtlm}
\end{mathletters}
Using (\ref{eq:trmtlm}) and (\ref{eq:rmlm}) to evaluate 
   $\widetilde{r_{m+1}}-\widetilde{l_{m}}$, 
one easily verifies that
\begin{equation}
   \epsilon
=
   \frac{\widetilde{r_{m+1}}-\widetilde{l_{m}}}{r_{m}-l_{m+1}} ,  
\label{eq:eps}
\end{equation}
is constant along the chain.  
This number fully characterizes the phase-space contraction 
of the multibaker map. 
In harmony with the common use in the dynamical systems literature, we 
will also say that $\epsilon$ characterizes the {\em dissipation.\/} 
It will be kept constant when taking the macroscopic limit. 

There are two special values of the parameter $\epsilon$ which are worth 
mentioning.  
When $\epsilon$ takes the value 
   $\epsilon = 1$, 
the phase-space dynamics is locally area preserving 
($\widetilde{r_{m+1}}=r_m, \widetilde{l_{m}}=l_{m+1}$).
For
   $\epsilon = -1$ 
($\widetilde{r_{m+1}}=l_{m+1}, \widetilde{l_{m}}=r_{m}$) 
we call the resulting phase-space dynamics time reversible, since the initial
area of any small region is recovered after taking an arbitrary closed path
along the chain (in points differring from the initial one, 
however, the area is in general different from the initial one).

\section{Entropies and their Time Evolution}
\label{sec:entropies}
%

\subsection{Gibbs entropy $S^{(G)}$}
\label{sec:Gibbs}
For the generalized multibaker map, the Gibbs entropy is defined
in terms of the phase-space density $\varrho(x,p)$ and the kinetic-energy
density $w(x,p)$ as
\begin{equation}
   S^{(G)}
=
    - k_B \int \rmd x \,\rmd p \; \vrho(x, p) \;
      \ln\frac{\vrho(x,p)}{\vrho^{\star}(w(x,p))} ,
\label{eq:SG}
\end{equation}
where $\vrho^{\star}(w)$ is a reference density, which depends on
$w$, and, through it, also on the phase-space coordinates 
($k_B$ denotes Boltzmann's constant).
We write
   $\varrho^{\star}(w)$ in the form
\begin{equation}
   \varrho^{\star}(w)
=
   \frac{\varrho^{\star}}{f(w(x,p))}
\label{eq:vrhostar} 
\end{equation}
where $\varrho^{\star}$ is a constant reference phase-space density, and $f$
is a dimensionless function. 
The actual form of $f(w)$ will be determined below by the requirement of
consistency with thermodynamics. 
At the moment we assume only that it is sufficiently smooth to expand it to
second order in its argument $w$. 

The Gibbs entropy $S_{m}^{(G)'}$ after one time step can be expressed by
making use of (\ref{eq:nm'}) and (\ref{eq:wm'}) for the three columns of cell
$m$: 
\end{multicols}
\widetext
\begin{eqnarray}
   S_{m}^{(G)'}
&=&
- k_B ab \; \bigg[
   (1 - r_m - l_m) \varrho_m
      \ln\left( \frac{\varrho_m}{\vrho^{\star}} f(w'_{m,s}) \right)
\non\\
&&
   + r_{m-1} \varrho_{m-1}
      \ln\left( \frac{r_{m-1}}{\widetilde{r_{m}}} \;
                \frac{\varrho_{m-1}}{\vrho^{\star}} \;
                f(w'_{m,r})
         \right)
   + l_{m+1} \varrho_{m+1}
      \ln\left( \frac{l_{m+1}}{\widetilde{l_{m}}} \;
                \frac{\varrho_{m+1}}{\vrho^{\star}} \;
                f(w'_{m,l})
         \right)
\bigg] .
\end{eqnarray}
After inserting the update (\ref{eq:rhoprime}) for the phase-space density,
substracting  
   $S_m^{(G)}=- k_B ab \varrho_m \ln (\varrho_m f(w_m)/\varrho^{\star})$, 
and rearranging terms, one finds 
\begin{eqnarray}
   \frac{{S_m^{(G)}}' - S_m^{(G)}}{a\tau}
& = &  - k_B \frac{b}{\tau}
\bigg[
   (\varrho_m^{'} - \varrho_m) \;
   \ln\left( \frac{\varrho_m}{\varrho^{\star}} f(w_m) \right)
  +\varrho_m^{'} \ln\left( \frac{ f(w_{m,s}^{'})}{ f(w_m) } \right)
\non\\
&&
\qquad +
   r_{m-1} \varrho_{m-1}  \ln\left( \frac{r_{m-1}}{\widetilde{r_m}} 
               \frac{\varrho_{m-1}}{\varrho_m}
                               \frac{f(w'_{m,r})}{f(w'_{m,s})}
                       \right)
+
   l_{m+1} \varrho_{m+1}  \ln\left(  \frac{l_{m+1}}{\widetilde{l_m}}
                                     \frac{\varrho_{m+1}}{\varrho_m}
                                     \frac{f(w'_{m,l})}{f(w'_{m,s})}
                       \right)
\bigg]
\non \\[2mm]
& \equiv &
 -  \frac{ j_m^{(s)} - j_{m-1}^{(s)} }{a}
 +  \Phi^{\rm (thermostat)}
 \equiv \Phi . 
\label{eq:SGdot}
\end{eqnarray}
This can be interpreted as a balance equation for the Gibbs entropy.
The temporal change of $S^{(G)}$ comprises two contributions:
the divergence of an entropy current
\begin{mathletters}
\begin{equation}
j_m^{(s)}
\equiv
 -  j_m  k_B \ln\left( \frac{\varrho_m}{\varrho^{\star}} f(w_m) \right)
 +       k_B \frac{ab l_{m+1}}{\tau} \varrho_{m+1}
      \ln\left( \frac{\varrho_{m+1}}{\varrho_m} \frac{f(w_{m+1})}{f(w_m)} \right)
\label{eq:js-mikro}
\end{equation}
and a flux into the thermostat
\begin{eqnarray}
\frac{\Phi^{\rm (thermostat)}}{k_B} 
& \equiv &
- \frac{b}{\tau} \left[
   \varrho'_m \ln\frac{f(w'_{m,s})}{f(w_m)}
\right. 
\nonumber\\
&&
\left. 
   + r_{m-1} \varrho_{m-1} \ln \left(
                               \frac{{r_{m-1} }}
                             {\widetilde{r_m}}
                               \frac{f(w'_{m,r}) f(w_m)}
                            {f(w'_{m,s}) f(w_{m-1})}       \right)   
   +
      l_{m+1} \varrho_{m+1} \ln \left( 
                               \frac{{l_{m+1} }}
                               {\widetilde{l_m}}
                               \frac{f(w'_{m,l}) f(w_m)}
                            {f(w'_{m,s}) f(w_{m+1})}       \right)
\right].
\label{eq:phith}
\end{eqnarray}
\end{mathletters}
\begin{multicols}{2}
\noindent 
In general, this decomposition is not unique.
It will turn out that $\Phi^{\rm (thermostat)}$ contains terms, which
can be combined to a divergence and hence transferred to the entropy
current. 
This freedom can only be removed in the macroscopic limit, 
where in physically relevant situations the splitting is 
unique. 
In any case, the form of the current is close to the
thermodynamic one (\ref{eq:js-macr}):
it contains a contribution
proportional to the current density $j$ and another term
which by its dependence on the function $f(w)$ characterizes the
local kinetic-energy gradient.

We identify the temporal change of the Gibbs entropy with the 
flux of the coarse-grained entropy. 
This is meaningful from an information theoretic point of view. 
After all, the Gibbs entropy characterizes the information encoded in 
the {\em microscopic\/} time evolution of a system. 
Consequently, changes of this entropy may only be due to an entropy 
current and to a coupling to the thermostat, 
i.e., to terms like those identified in Eq.~\eq{SGdot}. 

\subsection{Coarse-grained entropy S}
\label{sec:entropy}
The coarse-grained entropy $S_m$ of cell $m$ is defined
in an analogous way as the Gibbs entropy \eq{SG}, but using
now the cell density $\vrho_m$ and the cell's kinetic energy
density $w_m$
\begin{equation}
   S_m = - k_B ab \vrho_m \ln\left( \frac{\vrho_m}{\vrho^{\star}} \, f(w_m) \right).
\label{eq:Sm} 
\end{equation}
The coarse-grained entropy of cell $m$ after one time step is
\begin{equation}  \label{eq:Smprime}
   S'_m
=
   - k_B ab\vrho'_m \ln\left( \frac{\vrho'_m}{\vrho^{\star}} \, f(w'_m) \right) .
\end{equation}
It depends on the updated values of the coarse-grained quantities.  

In order to find the balance equation for the coarse-grained entropy, 
we make use of the argument of the previous subsection that the change 
in the Gibbs entropy may be interpreted as the macroscopic entropy 
flux. 
This allows us to rewrite the temporal change of the coarse-grained entropy as 
\begin{eqnarray}
  \frac{\Delta S_m}{ \tau}
& \equiv &
   \frac{S_m' - S_m}{ \tau}
\non\\
& = &
   \frac{{S_m^{(G)}}' - S_m^{(G)}}{\tau}
+
   \frac{( {S_m}'- {S_m^{(G)}}' )  -  ( S_m - S_m^{(G)} )}
        {\tau}                                              .
\label{eq:micbal}
\end{eqnarray}
The first term on the right-hand side is the entropy flux 
   $\Delta_e S_m / \tau$ 
through cell $m$, and the second one represents an irreversible entropy 
production $\Delta_i S_m / \tau$. 
Hence, Eq.~(\ref{eq:micbal}) constitutes a discrete entropy balance in the
form  
\begin{equation}
   \Delta S_m = \Delta_e S_m + \Delta_i S_m 
\label{eq:micro-balance}
\end{equation}
with
\begin{mathletters}
\begin{eqnarray}
   \Delta_e S_m & = & {S_m^{(G)}}' - S_m^{(G)},
\label{eq:DeltaeSm}
\\
   \Delta_i S_m & = & ( {S_m}' - {S_m^{(G)}}' ) - ( S_m - S_m^{(G)} ) .
\label{eq:DeltaiSm}
\end{eqnarray}
\end{mathletters}
In the information theoretic interpretation of
entropies, the difference
   $S_m - S_m^{(G)}$
measures the information of a microscopic system which cannot be resolved by
a coarse-grained description. 
Hence, $\Delta_i S_m$ is the increase per time unit $\tau$ of the information
which cannot be resolved when characterizing the state of the system by
coarse-grained densities. 
It is positive by construction, and except for a transient behaviour obtained
for certain initial conditions (which will not be considered here), it can
only increase. 

Note that the entropy 
   $S_m$ and the difference
   $S_m - S_m^{(G)}$
might depend in general on details of the coarse graining. 
This dependence drops out { in the macroscopic limit}, when calculating temporal changes { (cf.~App.~\ref{app:stability}). 
All quantities appearing in the entropy balance \eq{micro-balance} 
will turn out to be 
thermodynamically well-defined observables. }

As a last step, we discuss the explicit form of the the rate of irreversible
entropy production. 
The initial condition that the system is prepared with uniform densities in
every cell, implies 
   $S_m=S_m^{(G)}$.
The irreversible entropy change during one time step is therefore the 
difference between the Gibbs and the coarse-grained entropy taken  
after one time step 
   $\Delta_i S_m= S_m^{'}- S_m^{(G)'}$.
It can be split into two parts. 
An $f$-independent part, which comprises contributions due 
to the particle current, and another one, which is related to 
inhomogeneities in the kinetic-energy density, and hence in  
temperature. 
We write 
\begin{equation}
   \Delta_i S_m  = \Delta_i S_m^{\rm (particle)} + \Delta_i S_m^{\rm (heat)}
\end{equation}
where
\end{multicols}
\widetext
\begin{mathletters} 
\begin{eqnarray}
\Delta_i S_m^{\rm (particle)}
& = &
k_B a b \; \left\{
   -{\varrho_m}' \;
   \ln\frac{{\varrho_m}'}{\varrho_m} 
   + r_{m-1} \varrho_{m-1}  \,
      \ln\left[  \frac{r_{m-1}}{{\tilde r}_m} \frac{\varrho_{m-1}}{\varrho_m}  
         \right] 
   + l_{m+1} \varrho_{m+1} \,
      \ln\left[  \frac{l_{m+1}}{{\tilde l}_m} \frac{\varrho_{m+1}}{\varrho_m} 
         \right] 
\right\},
\label{eq:DiSm1}
\end{eqnarray}
and
\begin{eqnarray}
\Delta_i S_m^{\rm (heat)}
& = &
 k_B a b \; \left\{
   -{\varrho_m}' \;
   \ln \frac{f({w'_m})}{f(w'_{m,s})}
   + r_{m-1} \varrho_{m-1} \; \ln \frac{f\left(w'_{m,r} \right)}{f(w'_{m,s})}
   + l_{m+1} \varrho_{m+1} \; \ln \frac{f\left(w'_{m,l} \right)}{f(w'_{m,s})}
\right\} 
.
\label{eq:DiSm2}
\end{eqnarray}
\label{eq:DiSm}
\end{mathletters} 
\begin{multicols}{2}
\noindent 
All terms appearing in \eq{DiSm} have a proper physical meaning. 
The respective first ones characterize the change of entropy due to the time
evolution of the coarse-grained fields.  
The others amount to an entropy of mixing of regions in phase space with
different phase-space or kinetic-energy densities, respectively.  
Note that contributions from phase-space contraction appear 
only in $ \Delta_i S_m^{\rm (particle)}$.

\section{The Macroscopic Limit} 
\label{sec:limit}
%
\subsection{Definition of the limit}
\label{sec:def_limit}
The macroscopic limit implies that $a\ll L$, $N\gg 1$ and 
$\tau$ is much smaller than typical macroscopic time scales (for instance
$L^2/D$).  
Formally it is defined as
\beq
   a,\tau \rightarrow 0
\eeq
such that the spatial coordinate 
\begin{equation}
   x \equiv a m 
\end{equation} 
is finite. 
The phase-space density $\varrho_m$ integrated over the momentum-like variable
$p$ becomes the local particle density $n(x)$. 
Since $\varrho_m$ is independent of $p$, the integration corresponds to a
multiplication with the vertical cell size $b$. 
As mentioned earlier, the field $w$ is assumed to go over into the local
temperature $T(x)$ in the macroscopic limit. 
Thus, we have  
\begin{mathletters} 
\begin{eqnarray}
   b \varrho_m & \rightarrow & n(x),
\\
   w_m & \rightarrow & C T(x) ,
\end{eqnarray}
\end{mathletters} 
where $C$ is a constant of dimension one over temperature.
Moreover, the local drift, diffusion and the source strength 
$q$ are kept finite while taking the limit 
\begin{mathletters}
\begin{eqnarray}
   D    & \ra & D ,
\\
   v_m  & \ra & v(E,T,\partial_x T)
\\
   q_m  & \ra & q(x)   
\label{eq:vm}
\end{eqnarray}
\label{eq:Dvm}
\end{mathletters}
$E$ denotes the external field. 
In the following we do not write out the x-dependence of the fields, of the
drift and of the source term explicitly.

\subsection{Number and kinetic-energy density} 
We first notice that under the assumption of smoothness 
the spatial dependence of the two fields can be expressed
as 
\begin{eqnarray}
   b \varrho_{m\pm 1}
&\rightarrow &
   n \pm a \dx n  +  \frac{a^2}{2} \dx^{2} n , \non\\
   w_{m\pm 1}
&\rightarrow &
   CT \pm a \dx (CT)  +  \frac{a^2}{2} \dx^{2} (CT) .
\label{eq:n+-1}
\end{eqnarray}

In order to calculate the macroscopic limit of the particle current 
\eq{jm+b} 
we use \eq{rmlm} and \eq{Dvm}, to obtain 
\begin{equation}
   j = v n - D \dx n .
\label{eq:jbaker}
\end{equation}
{F}rom (\ref{eq:dt_varrho'}), the time evolution  of
the density $n$ can be obtained as:
\begin{equation}
\dt n
   =
-  \6_x (n v)
+ D \dx^{2} n.
\label{eq:nmprimeovernm}
\end{equation}

Similarly, we have for the time evolution of $n T$ 
[cf.~Eqs.~\eq{dt_varrho_m-w}, \eq{jmw+b}] 
\begin{equation}
\dt (nT)
   =  -
\dx j^{(nT)}  + n T q
\label{eq:nT}
\end{equation}
with the $nT$-current 
\begin{equation}
 j^{(nT)}
   =
 T j - n   D \; \dx T . 
\label{eq:jnT}
\end{equation}

\subsection{Irreversible entropy production}
\label{sec:production}
For the contribution \eq{DiSm1} of the particle current to the 
irreversible entropy production we obtain (cf.~the analogous
calculation in \cite{VTB98,BTV98} for details) 
\begin{mathletters}
\begin{equation}
   \frac{\Delta_i S_m^{\rm\tiny (particle)}}{k_B a \tau}
= \frac{j_{\epsilon}^2}{nD} ,
\label{eq:Ph1}
\end{equation}
where
\beq
j_{\epsilon} \equiv j-\frac{1+\eps}{2} \; n v
   = \frac{1-\epsilon}{2} \; n v - D \dx n 
\eeq
agrees with the particle current (\ref{eq:jbaker}) up to the  dissipation
dependent term $(1-\epsilon)/2$.  

The other contribution \eq{DiSm2} to the irreversible entropy production
comprises the explicit dependence on the kinetic-energy field.  
 can be evaluated in a straightforward manner by Taylor expanding the
function $f$ and the logarithms to quadratic order around $w^{'}_{m,s}$.  
The terms linear in $ f^{'}(w^{'}_{m,s})/f(w^{'}_{m,s}) $ exactly cancel. 
(Here $f'(z)$ denotes the derivative of $f(z)$ with respect to $z$). 
In nonvanishing order the macroscopic limit is therefore 
\begin{equation}
   \frac{\Delta_i S_m^{\rm\tiny (heat)}}{k_B a\tau}
 =
   n D {(\dx CT)}^2
\;
   \left[\frac{f''(CT)}{f(CT)} -
      \left( \frac{f'(CT)}{f(CT)}
      \right)^2
   \right].
\label{eq:Ph2}
\end{equation}
\end{mathletters}
Note that the square bracket can also be written as the second derivative of
$\ln{f}$. 

\subsection{Entropy flux}
\label{sec:js}
By expanding $f(w_{m\pm 1}$) to linear order around $w_m$, one finds for the
macroscopic limit of the entropy current \eq{js-mikro} 
\begin{mathletters} 
\begin{equation}
   \frac{j^{(s)}}{k_B} 
=
   - j \; \ln\frac{n f(CT)}{n^{\star}} 
   +     D \dx n
   +     n D \; \dx (CT) \;
   \frac{f'(CT)}{f(CT)} ,
\label{eq:js}
\end{equation}
where $n^* \equiv b \vrho^*$ is a reference particle density, which 
is constant in space and time. 

The macroscopic limit of the  entropy flux into the thermostat
(\ref{eq:phith}) is found to be 
\begin{eqnarray}
&   & 
\frac{\Phi^{\rm (thermostat)}}{k_B}
 = 
   - n CT 
   \frac{f'(CT)}{f(CT)}
   q
\nonumber
\\
&   & \qquad \qquad 
- \frac{v}{D}
\left[
   \frac{(1-\epsilon)^2}{4} n v + \epsilon D \dx n \right] 
   + \dx (n v) .
\label{eq:Phi}
\end{eqnarray}
\end{mathletters} 
It contains the spatial derivative $\dx (n v)$,
which underpins our earlier statement that the splitting of the entropy flux
into the divergence of a current and a flux going directly into a thermostat
is not unique.  
It is natural to remove the derivative from $\Phi^{\rm (thermostat)}$,  which
leads to the entropy current 
\begin{mathletters} 
\begin{equation}
   \frac{j_1^{(s)}}{k_B} 
=
   - j \left[1+  \ln\frac{n f(CT)}{n^{\star}} \right]
   +   n D \; \dx (CT) \;
   \frac{f'(CT)}{f(CT)} ,
\label{eq:js1}
\end{equation}
and to the flux 
\begin{eqnarray}
\frac{\Phi_1^{\rm (thermostat)}}{k_B} 
& = &
   - n CT 
   \frac{f'(CT)}{f(CT)}
   q
\nonumber\\ &&
- \frac{v}{D}
\left[
   \frac{(1-\epsilon)^2}{4} n v + \epsilon D \dx n
\right] .
\label{eq:Phi1}
\end{eqnarray}
\end{mathletters} 
However, there are other $\epsilon$-dependent splittings, too.
For instance, by adding $(1+\epsilon)nv/2$ to the entropy current one obtains 
\begin{mathletters} 
\begin{equation}
   \frac{j_2^{(s)}}{k_B} 
=
   - j \ln\frac{n f(CT)}{n^{\star}} 
   -   j_{\epsilon}
   +   n D \; \dx (CT) \;
   \frac{f'(CT)}{f(CT)} ,
\label{eq:js2}
\end{equation}
and
\begin{equation}
\frac{\Phi_2^{\rm (thermostat)}}{k_B}
 =
   - n CT 
   \frac{f'(CT)}{f(CT)}
   q
-  \frac{1-\epsilon}{2} \frac{v}{D} j_{\epsilon}
+  \frac{1+\epsilon}{2} n \dx v
.
\label{eq:Phi2}
\end{equation}
\end{mathletters} 
This shows that the splitting of the total entropy flux 
into a divergence  of an entropy current and a flux 
   $\Phi^{\rm (thermostat)}$ 
is in general not unique for arbitrary values of $\epsilon$,
not even in the macroscopic limit.  

\section{Consistency with thermodynamics}
\label{sec:compare}
Having found the general expressions for the macroscopic limit of the
particle and the entropy flux, and of the irreversible entropy production, we
are now in a position to make specific choices 
for the parameter $\epsilon$,
for the yet undetermined functions $q(x)$ and $f(w)$, and for the functional
$v(E,T,\partial_x T)$.  

Comparing (\ref{eq:jbaker}) with the thermodynamic particle current 
(\ref{eq:j1}), we find that the drift $v(E,T,\partial_x T)$ must 
take the form 
\begin{equation}
   v = v_{el} - k\frac{D}{T} {\dx T}.
\label{eq:v-def}
\end{equation}
Since earlier we have not found any neccessity to fix $v$,
this choice is obviously consistent with thermodyamics.
It remains however to be seen if the other constraints
can be fulfilled.

The form of $f$ can be fixed by observing that the term  
$(\dx T/T)^2$ appears in the irreversible entropy production 
(\ref{eq:sirr-therm}) with the {\em same} coefficient as the $-\dx T/T$ 
term in the entropy current (\ref{eq:js-macr}). 
Comparing the $f$-dependent parts in (\ref{eq:Ph2}) and 
(\ref{eq:js}) [or (\ref{eq:js1}) or (\ref{eq:js2})] we find that this 
can only happen if 
\begin{equation}
   z (\ln f(z))''= - (\ln f(z))'. 
\end{equation}
The solution of this differential equation is a power law  
\begin{equation}
   f(z) = z^{-\gamma}
\label{eq:gamma-def}
\end{equation}
with $\gamma$ as a free constant parameter.
A constant prefactor can be absorbed into the definition \eq{vrhostar}
of $\varrho^\star$. 

Concerning the value of $\epsilon$, there are several constraints, which all
lead to the same  unique choice.  
(i) The requirement to have the same coefficient in front of the $v_{el} n$
and the $- D \dx n$ terms in the entropy current (\ref{eq:jsthermo1}) fixes
the value of $\epsilon$ to be $-1$ [cf.\  $j^{(s)}_2$ given by
(\ref{eq:js2})]. 
(ii) A natural splitting of the entropy flux into the negative 
divergence of the entropy current and a flux into the thermostat holds 
for $\epsilon=-1$, too. 
In particular, we then have 
   $j^{(s)}_1 = j^{(s)}_2$ 
and 
   $\Phi^{\rm (thermostat)}_1 = \Phi^{\rm (thermostat)}_2$. 
(iii) 
{ 
The particle flux into the thermostat 
(\ref{eq:Phi1}) [or (\ref{eq:Phi2})] has a well-defined meaning only 
when its second term contains the full particle current $j$. 
For all these reasons the only dynamics  which leads to physically 
acceptible results corresponds to the choice $\epsilon=-1$, which was
connected to a time-reversible dissipation mechanism in
Sect.~\ref{sec:parametrize}. 
For this choice also the particle-current dependent part of the entropy production 
(\ref{eq:Ph1}) contains the full square of $j$ as required for
consistency with the corresponding thermodynamic 
contribution (\ref{eq:sirr-therm}). }

With these choices
\begin{equation}
   \frac{j^{(s)}}{k_B} 
=
   - j   \left[  1+ \ln\frac{n (CT)^{-\gamma}}{n^{\star}}  \right]
   -     n D \gamma \; \frac{\dx T}{T} \;   .   
\label{eq:js-limit}
\end{equation}
In view of (\ref{eq:js-macr}), we identify the thermal conductivity
$\lambda$ and the Peltier coefficient $\Pi$ as 
\begin{mathletters}
\begin{eqnarray}
   \lambda & = & k_B n D \gamma,
\label{eq:lambda}
\\
   \frac{\Pi}{T}
& = &
  \frac{k_B}{e} \left( -1 - \ln\left[ \frac{n}{n^{\star}} (CT)^{-\gamma} \right] \right) . 
\label{eq:Pi}
\end{eqnarray}
\end{mathletters}
By this, the transport coefficients could be expressed by system 
parameters. 
It is remarkable that finite values were found for $\lambda$ and $\Pi$
although only the finiteness of $v$ and $D$ was assumed in the course of the
macroscopic limit. 

Similarly, for the flux into the thermostat one obtains 
\begin{equation}
\frac{\Phi^{\rm (thermostat)}}{k_B}  
=
   \gamma n q - \frac{v j}{D} .
\end{equation}
It contains a term  $vj/D$ corresponding to the change of entropy associated
with Joule's heating due to the drift $v$ of the particles.  
In thermodynamics $\Phi^{\rm (thermostat)}=0$ in the bulk, so that the source
term $q$ takes the form  
\begin{equation}
   \frac{q^*}{k_B} = \frac{vj}{\lambda}.
\label{eq:q*}
\end{equation}
It describes the increase of the local kinetic energy due to
dissipative heating. 
The heat thus deposited in the system will be transported to the
boundaries by a heat current.

The entropy density obtained as the macroscopic limit of \Eq{Sm} is 
\begin{equation}
  s =
   - k_B n \ln\left[ \frac{n}{n^{\star}}{(CT)}^{-\gamma} \right] . 
\label{eq:entr}
\end{equation}
This implies that, up to an additive constant,  $e\Pi/T$ is
the entropy per particle $s/n$.  

Since the multibaker map describes a system of weakly interacting \cite{weak}
particles, it is natural to assume that not only its entropy function
\eq{entr} but also its chemical potential corresponds to that of a classical
ideal gas. 
We take
\begin{equation}
\mu_c \equiv 
   k_B T ( 1 + {\gamma})
     - T \frac{s}{n}.
\label{eq:muc}
\end{equation}
{F}rom these two equations of state $\gamma$ follows to be the specific heat at
constant volume, measured in units $k_B$.  

By substituting the chemical potential \eq{muc} into the thermodynamic
expression of the Peltier coefficient \eq{Pe} and comparing it with
the particular form obtained for $\Pi /T$ in \eq{Pi}, one immediately
sees that the coefficent $k$ of \eq{v-def} has to vanish. 
Hence, $\varsigma = e \alpha$ from \eq{k}, and from \eq{varsigma} 
one recovers the Onsager relation $\Pi=T \alpha$.  
The fact that $k$ turns out to be zero seems to be a special feature of the
baker model with three strips.   
In this version temperature cannot move without an explicit particle
motion, and hence no thermal diffusion is expected. 

Next, we consider the heat current $j^{(heat)}$, i.e., the energy current 
from which the potential energy of the external field is excluded 
\begin{equation}
   j^{(heat)}
 \equiv 
   Tj^{(s)}+\mu_c j .
\end{equation}
Inserting the explicit form of currents $j$ and 
$j^{(s)}$ [cf.~(\ref{eq:jbaker}) and (\ref{eq:js-limit})] into
\Eq{ju-def}, this yields  
\begin{equation}
   j^{(heat)}
   =
   -\lambda \6_xT + k_B T\, \gamma j.
\label{eq:ju'}
\end{equation}
This form is indeed consistent with thermodynamics, and also with
the macroscopic limit of the $nT$-current (\ref{eq:jnT}). 
Multiplying the latter by $k_B \gamma$, we recover (\ref{eq:ju'}). 

It is worth poiting out that due to the definition
\eq{Diffusion} of the diffusion coefficient and the particular form
of the chemical potential \eq{muc}, we find that
\begin{equation}
   \sigma_{el} k_B T = e^2 D n. 
\label{eq:Einstein}
\end{equation}
Consequently, Einstein's relation holds in the multibaker map.
One can thus express the electric 
conductivity by the diffusion coefficient $D$ in all formulas. 
In particular, we find 
\begin{equation}
   \sigma^{\rm (irr)}
=
   \lambda \left( \frac{\nabla T}{T} \right)^2
 + \frac{k_B j^2}{nD},
\label{eq:sirr}
\end{equation}
a formula which in the case of constant temperature has already been derived
in earlier versions of the multibaker map \cite{VTB98,BTV98}. 

As further consequences of Einstein's relation, we mention that:
(i) the electric drift $v_{el}$ \eq{vs} is proportional to $E/T$: 
\begin{equation}
   v_{el}=\frac{eD}{k_B}\frac{E}{T} 
\end{equation}
since the diffusion coefficient
is assumed to be constant.
(ii) comparing heat and the electric conductivities \eq{lambda} and 
\eq{Einstein}, we find 
\begin{equation}
   \frac{\lambda}{\sigma_{el} T} 
= 
   \gamma \left( \frac{k_B}{e} \right)^2 , 
\end{equation} 
which implies that this ratio is independent of thermodynamical state
variables.  
Thus, the Wiedermann-Franz law \cite{AM} proves to hold for the multibaker
model.  
(iii) the elementary Drude theory \cite{AM} of
metallic conduction predicts the Seebeck coefficient $\alpha=\Pi /T$ to be
proportional to $k_B/e$, i.e., to be independent of temperature or density,
which contradicts observation.  
Such a term is indeed present in \Eq{Pi}, but its second term also predicts a
specific state dependence.  
Thus, the present model turns out to describe certain features of transport
more realistically than the classical Drude model, 
although it cannot be expected 
to give a microscopically realistic theory of thermoelectric phenomena, 
(which contain essential quantum effects due to the strong degeneracy of the 
fermionic electron gas) \cite{Boltzmann}. 

Finally, we consider the temperature equation following from
(\ref{eq:nT}).  
It takes the form 
\begin{equation}
   \dt T = q T + \frac{\dx (\lambda \dx T)}{k_B \gamma n} - j \frac{\dx T}{n} ,
\label{eq:T}
\end{equation}
which can be shown to be 
consistent with the general relation (cf. XIII. (85) of \cite{GM})
expressing the entropy's local  
time derivative as 
\begin{equation} \label{eq:sbalance}
   \dt s 
= 
   \frac{ \dx (\lambda \dx T)}{T} 
-  \dx \left(\frac{e \Pi j}{T} \right) 
+ \frac{e j^2}{\sigma_{el} T} . 
\end{equation}
Here, the respective  
terms accounts for { the} heat conduction, { the} Peltier and 
{ the} Joule heating.  
Substitute (\ref{eq:entr}) and the Peltier coefficient, \Eq{Pi}, to see that
\Eq{sbalance} becomes an identity if and only if the thermodynamical choice of 
\Eq{q*}, the source term, is taken.  

\section{Thermostatting}
\label{sec:therm}
After having identified the condition for full consistency with 
thermodynamics in the form of $\Phi^{\rm (thermostat)}=0$ or $q=q^*$, 
we turn to a short discussion of cases where there can be
an entropy flux into the thermostat. 
In the thermostatting algorithm of non-equilibrium molecular 
dynamics \cite{books}, heat is taken out of the system in order to keep 
the temperature constant in a spatially homogeneous steady state, and to 
avoid an overheating due to the permanent acceleration produced by en 
electric field. 
In our setting this corresponds to a case with $\dx T=0$. 
Such a uniform temperature field is stationary for 
$q=0$ only, as follows from the temperature equation (\ref{eq:T}), 
so that 
   $\Phi^{\rm (thermostat)}=- k_B v j /D$.
It is indeed a kind of Joule's heat, which is let into the 
thermostat. 
Note, however, that classical thermodynamics does not admit a 
stationary homogeneous state $\dx T=0$ to be steady, since the 
temperature increases in the bulk due to Joule's heating. 
This indicates that thermostatting is a tool by which one can turn a
preselected temperature profile into a steady state. 
After all, for every density profile consistent with given boundary conditions
and a preselected fixed temperature profile $T(x)$ there is a source term
distribution $q(x)$, such that the temperature does not change in time
[cf.~(\ref{eq:T}) with $\dt T=0$]. 
In all these cases 
   $\Phi^{\rm (thermostat)}=k_B \left( \gamma n q- v j /D \right) $ 
is different from zero. 
This shows that the algorithm for thermostatting can be maintained even for
complicated temperature profiles. 
Since the source $q$ appears neither in the currents, nor in the 
irreversible entropy production, 
nor in the constraint for the Onsager relation 
to hold, nor in the transport  coefficients, the local entropy balance 
is consistent with every choice of $q$ or $\Phi^{\rm (thermostat)}$ 
provided the entropy flux appears in the form of \eq{Phi-thermo}. 
In general, the divergence of the entropy current $j^{(s)}$ contributes to the
entropy flux $\Phi$, and 
the remaining part of the reversible change of the entropy is transferred to
the thermostat. 

We conclude that although thermostatting is a deviation from 
classical thermodyanamics, it seems to be the {\em weakest\/} possible 
deviation in the sense that except for the form of the entropy flux it
leaves all {\em local\/} thermodynamic relations  invariant. 
It can thus be seen as 
an idealization of a physical thermostat (which can in reality only be
attached to the boundaries of a system), 
where heat need not be transported spatially (to the boundaries), but
can directly be released into the surroundings. 
This gives rise to a non-thermodynamic contribution to the entropy flux in 
the local entropy-balance equation in a generalization of 
non-equilibrium thermodynamics [cf.~Eqs.~\eq{Phi-thermo} and \eq{dts}]. 
This  
generalization of the local balance equation, 
however, does not imply at all a similarity of {\em global\/} 
{ transport properties, which in general depend on the spatial
dependence of the fields.}
It is clear from (\ref{eq:T}) that a state which is steady with a given 
$q(x)$ will not be steady with the thermodynamic choice $q^*(x)$. 
It will not even have similar density profiles. 
{ Therefore,} thermodynamics and thermostatted 
descriptions might lead to strongly different results 
{ on the global level.}  

\section{Discussion}
\label{sec:conclude}
We have extended multibaker models by augmenting the density field $\varrho$
of these models by a temperature-like field variable $w$. 
This allowed us to address 
problems like thermoelectric
cross effects requiring two independend thermodynamic driving fields.
The model has the following features:

(A) The evolution equation of $w$ requires source terms reflecting the
local irreversible heating of the system in the presence of
transport.

(B) The temperature enters the entropy through a 
kinetic-energy dependent normalization of the (phase-space) density.

(C) Consistency with the thermodynamic description of transport is achieved
for densities which are coarse grained in regions of small 
spatial extension. 

(D) Comparing the coarse-grained description with the microscopic one,  
allows us to identify all
contributions to the {\em local\/} entropy balance. 

(E) The time evolution of the system can be interpreted as that of 
weakly interacting particles, whose motion may only be coupled 
through a mean-field like dependence of the evolution equations on the
coarse-grained field variables. 
In accordance with this, the resulting ``multi-baker'' gas 
obeys the classical ideal-gas equation of state. 
The Onsager relation, the Wiedemann-Fanz law and the Einstein
relation can be derived, and expressions are found for the Peltier and Seebeck
coefficients. 

(F) The local entropy balance of non-equilibrium thermodynamics
can be generalized by introducing at every location an instantaneous 
flow of entropy (i.e., of heat) into a thermostat. 
When time reversibility is maintained, the dynamics becomes closely
reminiscent to numerical algorithms related to Gaussian thermostats. 

{ 
(G) 
Dissipation and thermostatting play different 
roles in dynamical-system models for transport. 
The condition $\epsilon=-1$, for time reversibility, was found
independently of the choice of $\Phi^{\rm (thermostat)}$. 
With this dissipation we can describe both thermostatted and 
non-thermostatted systems. 
} 

It is remarkable that an agreement with thermodynamics
could be achieved by this comparatively simple model. 
Indeed, strong restrictions on the choice of its parameters were needed. 
The free functions (drift $v$, source strenght of heat $q$ and
normalization of densities $f$) had to be chosen appropriately,
and a free dissipation-related  parameter ($\epsilon$) had to be fixed
to a given value leading to a time-reversible dissipative dynamics
(even a Hamiltonian, volume-preserving dynamics is excluded).
With the given choices, however, we do not find any restriction to weak
gradients.
This might be a consequence of the baker map's piecewise linear
character and { of} the Markov properties resulting from the fact
that this family of maps admits no pruning. 

We have used a generalized concept of dynamical systems to model 
open boundaries where transport can be induced by appropriately chosen
boundary conditions. 
The time evolution of a macroscopically large number of independent `particles'
is considered. 
Consequently, not even in 
steady states the natural measure of the
multibaker map is relevant to calculate physical
observables. 
After all, this measure is only defined if the map is closed by periodic
boundary conditions. 
Rather another measure, the one {\em forced\/} on the system by the open 
boundary conditions, plays the central role. 
Such measures were first investigated by Gaspard and coworkers
\cite{Gasp,G,TG}. 

Finally, we draw attention to the fact that the present model differs
in important features from other models of transport by low-dimensional
dynamical systems. 
The transition probabilities (which are closely related to the drift and
diffusion coefficients) may depend on the coarse-grained fields.  
This dependence 
leads to a dynamical system with many degress of freedom. 
Consequently, the full time evolution of the $(x,p)$ dynamics together with
the time dependence of the parameters can be interpreted as a peculiar coupled 
map lattice designed to closely follow transport equations. 
In a steady state, however, the parameters 
take time-independent values such that the time evolution of the
particles is described by a 
two-dimensional dynamical system, 
a map acting on $(x,p)$, which is in general lacking translation
invariance.  

It is worth emphasizing that a further reduction of the dimension is
impossible. 
By neglecting the $p$ variable (i.e., when projecting the baker map to obtain
a one-dimensional map describing the tranport of particles along the
$x$-direction) one finds full consistency with macroscopic transport 
equantions, but all drift-dependent terms disappear
from the entropy balance, 
which therefore deviates from its thermodynamic form 
(\ref{eq:dts},\ref{eq:sirr-Phi-therm}). 
Hence, modelling only the transport processes via dynamical systems is a
much easier enterprise than aiming also for a proper description of
entropies. 
{F}rom the point of view of a correct entropy balance the
existence of a phase-space variable orthogonal to the transport direction is
essential. 
Only in this case can the fractal structures in the microscopic
densities be followed, whose unresolvability leads to entropy
production.  
It is open at present, however, if a dissipative dynamics is needed 
for this, since  
a variation of the cell size in the sense of \cite{TG98} 
might { convert contributions to the entropy production  
due to local phase-space contraction into those of mixing. } 

The suggested method for modelling thermoelectric cross effects
can be considered as a combination of a dynamical-system and a hydrodynamic
description. 
{ Besides the appearance of a source term of the kinetic energy, } 
the strong {\em mixing\/} character of the 
{ chaotic dynamics is essential,} 
which leads to
fractal phase-space patterns in the considered forced measure. 
By that it ensures that irreversibility, and thus consistency with
thermodynamics, can be reached in a description based on a 
low-dimensional 
dynamical system.

\acknowledgements
We are grateful to G.\ Nicolis, J. R. Dorfman, B. Fogarassy, 
J. Hajdu, G. Tichy and H. Posch for enlightening
discussions, and to the International Erwin Schr\"odinger Institute,
Vienna, for its kind hospitality.
Support from the Hungarian Science Foundation (OTKA T17493,
T19483), and the TMR-network, contract no. ERBFMBICT96-1193, is
acknowledged.

\end{multicols} 
\widetext 
\appendix 

\section{Rigorous definition of the map and of  the time evolution
of densities}
\label{app:definition}  
Every point $(x, p)$ with 
   $x \in [0,aN]$, 
   $1 \leq m \leq N$, and 
   $y \in [0,b]$ 
is mapped by the multibaker map $B$ as follows 
\begin{equation}
   B(x, p) 
= 
\left\{
\begin{array}{lllrcl}
  l_m^{-1} x + a (m-1), 
& \widetilde{l_{m-1}} p 
&  \hbox{\ \ for} & 0 &<(x-am)/a< & l_m 
\\    
  s_m^{-1} [x-a l_m]+am, 
& b \widetilde{l_m}+ \widetilde{s_{m}} p,   
&  \hbox{\ \ for} & l_m &<(x-am)/a< & l_m + s_m
\\
  r_m^{-1} [x-a(l_m+s_m)]+a(m+1),  
& b(\widetilde{l_{m+1}}+\widetilde{s_{m+1}})+\widetilde{r_{m+1}} p , 
&  \hbox{\ \ for} & l_m + s_m &<(x-am)/a< & 1 
\end{array} 
\right.
\label{eq:bakermap}
\end{equation}
\begin{multicols}{2} 
%
The dynamics of the phase-space density $\varrho (x,p;t)$ 
can be interpreted as the time evolution of a typical set of 
points distributed in the phase space. 
It is governed by the Frobenius-Perron equation 
\begin{equation}
   \varrho(z; {t+\tau})= \int \rmd z' \; \delta (z-B(z')) \varrho(z'; t) ,
\label{eq:FP}
\end{equation} 
where $z$ denotes points $(x,p)$ in the phase space of the multibaker chain. 

The boundary conditions are that $\varrho(z;t)$ is fixed to 
   $\varrho_0 $ and $\varrho_{N+1}$ 
in the cells $m=0$ and $m=N+1$, respectively. 
They are taken into account by the equation 
\begin{equation}
\varrho(z; {t+\tau})= \varrho_{0(N+1)} 
                     \int  \rmd z' \; \delta (z-B(z')) 
\end{equation} 
for points $z$ whose preimages lie in cell $0$ ($N+1$). 

Due to the chaoticity of the map and the difference in the boundary 
conditions, the density $\varrho(z;t)$ becomes more and more 
irregular as time goes on (at least for any smooth initial distribution). 
Therefore, asymptotically the concept of density is not 
well defined.  
For $t \rightarrow \infty$ one assigns the measure 
   $\mu (A) \equiv \lim_{t\rightarrow\infty}\int_A \rmd z\, \varrho(z;t)$ 
to any phase-space region $A$. 
 
Similarly to the phase-space density $\varrho(z; t)$, the
time evolution of the kinetic-energy density   
\begin{equation}
   \omega \equiv \varrho w,  
\end{equation}
is described by the integral equation 
\begin{equation}
   \omega(z; {t+\tau})
=
   (1+\tau q(z)) \; \int \rmd z' \; \delta (z-B(z')) \omega(z'; t). 
\end{equation}
In generalization of 
the Frobenius-Perron equation (\ref{eq:FP}), however, a source
term $q(x)$ is included now, 
which is 
is piecevise constant in the cells. 
The corresponding boundary conditions for $w(x,p;t)$ are 
\begin{equation} 
   \omega(x,p;t+\tau)=\varrho_{0(N+1)} w_{0(N+1)} \, [1+\tau q(x,p)] 
\end{equation} 
for points whose preimages are in cell $0$ or $N+1$, respectively.

Also in this case, $w(x,p; t)$ is no longer well-defined asymptotically. 
Instead, the stationary kinetic-energy distribution should be
considered as an invariant measure $\nu$, 
different from $\mu$, which assigns the weight 
   $\nu(A) \equiv \lim_{t\rightarrow\infty}\int_A \rmd z\, \omega(z;t)$ 
to every region $A$ in phase space.

\begin{figure}
\hspace*{2mm}                      
\epsfbox{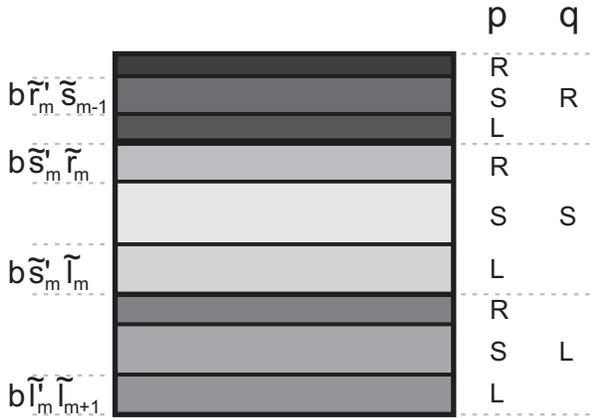} 
\vspace*{2mm} 
\caption[]{\narrowtext
Notations used to define the level-2 partitioning of a cell after two
applications of the map. 
Every strip is labeled by a two-letter sequence $p,q$, where 
$p,q \in \{L,S,R\}$ label the level-2 strips. 
($p$ stands for the level-2 strips within 
level-1 strips selected by the label $q$).  
To illustrate the use of these labels, the heights of a few strips are
indicated at the left side of the cell.
\label{fig:2step}}
\end{figure}
\section{Structural stability of macroscopic results}
\label{app:stability}

In this appendix we adapt the arguments of \cite{VTB98} to demonstrate
that, 
after taking the macroscopic limit,  
our results are independent of the detailed prescription of coarse graining. 
One part of such a demonstration should be that the same results are obtained 
when coarse-graining is applied to any number $m_c$ of 
successive cells after every time step. 
In this case all previous results for quantities of unit volume 
remain unchanged in the macroscopic limit for any fixed 
$m_c$, since both the length of the investigated region 
and the differences of the values of the coarse-grained  
fields become multiplied by a factor $m_c$, 
and thus the macroscopic gradients do not change. 

We therefore concentrate on the more involved 
case of coarse graining on the cells only after 
every $n$ time steps. 
From a density,   
which is constant in each cell, the dynamics generates 
structure on the $n$th level of the refining partition 
obtained by the $n$th images of the cells. 
Structural stability is demonstrated by 
showing that the total entropy production 
can be decomposed into a sum of contributions 
from coarse-graining on the different levels 
(starting with the finest structures) 
and that all these contributions are the same in the macroscopic limit. 
This is equivalent to showing that coarse graining may also 
to be applied to a finer partitioning within the cells, 
without affecting the resulting  thermodynamic relations. 
For sake of conciseness we immediately explore the physically relevant
case  
   $f(z) = z^{-\gamma}$ 
[cf.~Eq.~(\ref{eq:gamma-def})], 
suppress the normalization constant $\varrho^\star$ of the density  
and Boltzmann's constant $k_B$, 
and we only work out the case of $n$=2. 
Although this corresponds to just the simplest possibility, it 
indicates the strategy to be followed when discussing further refinements
\cite{refine}.

\subsection{Preliminaries} 

We first extend our notations in order to describe densities, entropies 
and fluxes defined on the different levels of coarse graining. 
To this end we consider the level-2 partitioning
(cf.~Fig.~\ref{fig:2step}) of the cells two time steps after the
initial coarse graining. 
Again primes denote quantities evaluated after the first application
of the mapping.  
Similarly, double primes are used for the values after two iterations. 
Note that also the transition probabilites $l_m$, $s_m$ and $r_m$ and
the widths and heights of the corresponding columns and strips receive
primes now, due to their possible dependence on $\varrho_m$ and $w_m$
(which evolve in time). 

In every cell $m$ there are $3^2$ strips labeled by a pair of 
symbols $(p,q)$ with $p,q\in\{L,S,R\}$; 
$p$ indicates the  
strip $(L,S,R)$ of Fig. \ref{fig-mbaker}b the point is located in 
after one time step 
(irrespective of the cell index), and $q$ specifies its position in the strip 
$(L,S,R)$ after the second time step. 
The strips $(p,q)$ have height $b\tilde{s}_{p,q}=b\tsmq\tsmqp$, 
and carry densities 
$\vrho''_m (p,q)$ and $w''_m (p,q)$. 
Here,  
\begin{equation} 
\tilde{s}_{m,q}
=
\cases{ 
       \tilde{r}_m,& for $q=R$  \cr
       \tilde{s}_m,& for $q=S$  \cr 
       \tilde{l}_m,& for $q=L$ }
\label{eq:tsmq} 
\end{equation}
is a shorthand notation for the height after one time step, and the subscript 
\begin{equation} 
m_q
=
\cases{
    m+1, & for $q=L$ \cr 
    m,  & for $q=S$ \cr
    m-1, & for $q=R$ }
\label{eq:m_q} 
\end{equation}
is used as a book-keeping device to indicate the cell where the points in a
given strip come from.  
An analogous definition holds for the quantities $s_{m,q}$ 
without tildes. 

We call 
   $\vrho''_m(p,q)$, $\vrho''_m(q)$, and $\vrho''_m$ 
   [$ \vrho''_m(p,q) w''_m(p,q)$, $\vrho''_m(q)w''_m(q)$, 
   and $\vrho''_m w''_m$] 
the level-2, level-1 and level-0 densities [kinetic-energy densities] 
after two time steps, respectively.   
By construction, the level-0 densities coincide with the cell densities
considered in preceding sections.  
The densities defined on different levels are related to each other since
coarse-graining preserves their average
value in each cell: 
\begin{mathletters} 
\begin{eqnarray}
\vrho''_m
     & = &
        \sum_q  \tsmq \vrho''_m (q)
      =
        \sum_q \tsmq  \sum_p \tsmqp \vrho''_m (p,q) , 
\\ 
\vrho''_m w''_m
     & = &
        \sum_q  \tsmq \vrho''_m (q) w''_m(q)
\nonumber\\
     & = &
        \sum_q  \tsmq \sum_p \tsmqp \vrho''_m (p,q) w''_m(p,q)  . 
\end{eqnarray} 
\label{eq:appRho} 
\end{mathletters} 

Based on the time evolution of the fields, relations between densities
at different instants of time can be calculated. 
Taking into account the respective action \eq{nm'} and \eq{wm'} of the mapping
on the densities $\varrho$ and $w$, we obtain 
\begin{mathletters} 
\begin{eqnarray}  
   \vrho_m^{''}(p,q)
&=&
   \frac{\smq}{\tsmq} \vrho_{m_q}^{'}(p) 
=
   \frac{\smq}{\tsmq} \frac{\smqp}{\tsmqp} \vrho_{m_{q,p}} ,  
\\ 
   w_m^{''}(p,q)
&=&
   w_{m_q}^{'}(p)(1+\tau {q_m}' ) 
\non \\ 
&=& 
   w_{m_{q,p}} (1+\tau q_{m_q}) (1+\tau {q_m}' )   . 
\end{eqnarray}    
\label{eq:vrhoTpp} 
\end{mathletters} 
Here $m_{p,q}$ denotes the action of \eq{m_q} applied to $m_p$: 
$m_{p,q}={(m_p)}_q$.
In order to follow the time evolution of the entropies, we consider
coarse-grained entropies defined with respect to densities of different
resolution. 
The level-$i$ entropy $S_m^{(i)}$ is defined with respect to the level-$i$
densities. 
Thus, 
   $S_m^{(0)} \equiv S_m = -ab\, \varrho_m \ln{(\varrho_m w_m^{-\gamma})}$, 
while the level-one and level-two entropies at time 
$\tau$ and $2\tau$ take the form
\begin{mathletters} 
\begin{equation}  
   S_m^{(1)'}
 =  
   -ab\,\sum_{p} \tilde{s}_{m,p} \vrho'_m (p)
      \ln \left[ \vrho'_m (p) {w'_m}^{-\gamma}(p) \right],
\label{eq:Sm1} 
\end{equation}
and
\begin{equation} 
   S_m^{(2)''}
 =  
   -ab\, \sum_{p,q} \tsmq \tsmqp \vrho''_m (p,q) 
      \ln \left[ \vrho''_m (p,q) {w''_m}^{-\gamma}(p,q) \right], 
\label{eq:Sm2}
\end{equation}
respectively.  
\end{mathletters}

\subsection{Entropy balance on level-1 strips} 

In order to obtain the entropy balance on level-1 strips, we relate
${S_m^{(2)}}''$ to ${S_m^{(1)}}'$. 
Making use of (\ref{eq:tsmq}), (\ref{eq:m_q}), (\ref{eq:vrhoTpp}) and
(\ref{eq:Sm2}), the level-2 entropy after two time steps $S_m^{(2)''}$
of cell $m$ can be worked out as 
\end{multicols} 
\widetext 
\begin{eqnarray}   
S_m^{(2)''}
&=&
   -ab\, \sum_{p,q} \tsmq \tsmqp \vrho_m^{''}(p,q) 
   \ln [ \vrho_m^{''}(p,q) {w^{'' \; -\gamma}_m(p,q)} ] 
\non \\
&=&
   -ab\, \sum_q \smq \vrmqp 
   \ln \frac{\smq}{\tsmq} 
-
   ab\, \sum_{p,q} \smq \smqp \vrho_{m_{q,p}}  
   \ln [ \vrmqpp \Tmqpp^{-\gamma} {(1+\tau {q_m}')}^{-\gamma} ]
\end{eqnarray} 
where
$\vrmqp=\sum_p \tsmqp \vrmqpp$.

By using (\ref{eq:Sm1}), $S_m^{(2)''}$ takes the form
\begin{equation}  \label{eq:Sm2pp}
   S_m^{(2)''}
=
    -
     ab\, \sum_q \smq \vrmqp \ln \frac{\smq}{\tsmq} 
    +
     \sum_q \smq S_{m_q}^{(1)'} 
\non\\
   +
    ab\, \gamma \sum_q \smq  \vrmqp \ln (1+\tau {q_m}') , 
\end{equation}
which, after carrying out the summation over $q$, leads  
to the level-one generalization of the the entropy flux 
\begin{eqnarray}   \label{eq:DeSm1p}
\Delta_e S_m^{(1)'}
\equiv 
   S_m^{(2)''} - S_m^{(1)'} = 
&-&
   ab\, r'_{m-1}
   \sum_p \tilde{s}_{m-1,p} \vrho_{m-1}^{'}(p) 
   \ln 
   \left(
   \vrho_{m-1}^{'}(p) w_{m-1}^{' \; -\gamma}(p) \frac{r_{m-1}}{\tilde{r}_m}
   \right) 
 \non \\
&-&
   ab\, l'_{m+1}
   \sum_p \tilde{s}_{m+1,p} \vrho_{m+1}^{'}(p) 
   \ln
   \left(
   \vrho_{m+1}^{'}(p) w_{m+1}^{' \; -\gamma}(p) \frac{l_{m+1}}{\tilde{l}_m} 
   \right) 
\non\\
&+&
   ab\, (r'_m+l'_m) \sum_p \tilde{s}_{m,p} \vrho_m^{'}(p) 
   \ln  
   \left(
   \vrho_m^{'}(p) w_m^{' \; -\gamma}(p)
   \right) 
+
   ab\, \gamma \sum_q \smq \vrmqp \ln (1+\tau {q_m}').
\end{eqnarray}
\begin{multicols}{2}
The entropy balance for the level-1 strips reads then 
\begin{equation}  
   \Delta S_m^{(1)'}
\equiv 
   S_m^{(1)''} - S_m^{(1)'} 
\equiv 
   \Delta_i S_m^{(1)'} + \Delta_e S_m^{(1)'} 
\end{equation} 
with the level-one irreversible entropy production 
\begin{equation} 
   \Delta_i S_m^{(1)'} = S_m^{(1)''} - S_m^{(2)''} .  
\end{equation}  
Again it is related to the loss of information on the microscopic 
state of the system when applying a coarse-grained description. 

If the densities $\varrho_{m_q}^{'}$ are uniform in the respective 
cells $m_q$, i.e., they do not depend on the partitioning label $p$, 
the above relations coincide with Eqs.~\eq{SGdot} and \eq{DiSm} [in which
$\vrho_m^{'}$ is eliminated via (\ref{eq:rhoprime})]. 
Typically, however, there is a non-vanishing difference 
\begin{equation} 
    \delta \Delta_i S_m^{(1)'} 
\equiv 
    \Delta_i S_m^{(1)'} - \Delta_i S_m^{(0)} , 
\end{equation} 
which only disappears in the macroscopic limit, as shown below after the
discussion of the entropy balance for coarse-graining after every second time
step.

\subsection{Entropy balance for coarse-graining after every second time step} 

In order to denote temporal changes taken with a time lag $2\tau$, we
assign the superscript $^{(2)}$ to $\Delta$. 
A direct consequence of (\ref{eq:DeSm1p}) is that the entropy flux
$\Delta_e^{(2)} S_m^{(0)}$ after two time steps is 
\begin{eqnarray}
\Delta_e^{(2)} S_m^{(0)}
&\equiv&
-S_m^{(0)}+S_m^{(2)''} \non\\
&=&
-S_m^{(0)}+S_m^{(1)'}-S_m^{(1)'}+S_m^{(2)''} \non\\
&\equiv&
\Delta_e S_m^{(0)}+\Delta_e S_m^{(1)'}. 
\end{eqnarray}
This flux is the two-time-step generalization of the entropy flux 
$\Delta_e S_m$. 

In order to establish the entropy balance, the change of the coarse-grained
entropy $\Delta^{(2)} S_m^{(0)}$ is considered (without coarse graining after
the first step) 
\begin{eqnarray}
   \Delta^{(2)} S_m^{(0)}
& \equiv &
   S_m^{(0)''}-S_m^{(0)}
=
   S_m^{(0)''}-S_m^{(2)''}+S_m^{(2)''}-S_m^{(0)} \non\\
&=&
   S_m^{(0)''}-S_m^{(2)''}+\Delta_e^{(2)} S_m^{(0)} .
\end{eqnarray}
Thus, the two-step irreversible entropy change $\Delta_i^{(2)} S_m^{(0)}$ 
takes the form
\begin{equation}  \label{eq:Di2Sm0}
   \Delta_i^{(2)} S_m^{(0)} 
\equiv
   \Delta^{(2)} S_m^{(0)} - \Delta_e^{(2)} S_m^{(0)}
=
   S_m^{(0)''}-S_m^{(2)''}.
\end{equation}
Again it is identified as the loss of information caused by coarse graining.  

Note that this rate of entropy production can be expressed as
\begin{equation}
   \Delta_i^{(2)} S_m^{(0)}
=
   \Delta_i S_m^{(0)'} + \Delta_i S_m^{(1)'}
\end{equation}
where 
$\Delta_i S_m^{(j)}= S_m^{(j)'}- S_m^{(j+1)'}$ $(j=0,1)$.
When coarse graining is applied after each time step, 
the entropy production is 
  $\Delta_i S_m^{(0)} + \Delta_i S_m^{(0)'}$. 
Thus, 
   $\delta \Delta_i S_m^{(1)'}$ 
also amounts to the difference between the irreversible entropy
production of the cases, where coarse graining is applied after each
and after every second time step, respectively.

\subsection{Evaluation of $\delta \Delta_i S_m^{(1)'}$} 

We recall that $\vrho_m^{''} (q)$ and $w_m^{''} (q)$ denote the fields after 
two time steps, when coarse graining is applied on the level-one partition. 
The level-one entropy after two time steps is 
\end{multicols} 
\widetext 
\begin{eqnarray}
S_m^{(1)''}
&=&
-ab\, \sum_q \tsmq \vrho_m^{''}(q) 
\ln \left( \vrho_m^{''}(q) {w_m^{''}(q)}^{-\gamma} \right) 
\non\\  
&=&
-
ab\, \sum_q \smq \vrmqp \ln \frac{\smq}{\tsmq}  
+   
\sum_q \smq \left[ -ab\, \vrmqp 
\ln \left( \vrmqp w_{m_q}^{'-\gamma} {(1+\tau q_m)}^{-\gamma} \right) \right] .
\end{eqnarray}
Using the form (\ref{eq:Smprime}) of the coarse-grained entropy, we obtain 
\begin{eqnarray}
   S_m^{(1)''}
&=&
  - ab\, \sum_q \smq \vrmqp \ln \frac{\smq}{\tsmq}
  + \sum_q  \smq S_{m_q}^{(0)'}  
+
  ab\, \gamma \sum_q \smq \vrmqp \ln(1+\tau q_m) 
\end{eqnarray} 
where $S_{m_q}^{(0)'}$ is the coarse-grained entropy of cell $m_q$
evaluated after one time step.  
This form of level-1 entropy can straightforwardly be compared with
the level-2 entropy (\ref{eq:Sm2pp}), leading to 
\begin{eqnarray}
   \Delta_i S_m^{(1)'}
&=&
   S_m^{(1)''}-S_m^{(2)''}   
= 
   \sum_q \smq [S_{m_q}^{(0)'}-S_{m_q}^{(1)'}]  
\equiv 
    s_m\Delta_i S_m^{(0)} 
   +r_{m-1} \Delta_i S_{m-1}^{(0)}
   +l_{m+1} \Delta_i S_{m+1}^{(0)}   
\non\\
&=&
    \Delta_i S_m^{(0)} 
-
    r_m [\Delta_i S_m^{(0)}-\Delta_i S_{m-1}^{(0)}]
    -l_m [\Delta_i S_m^{(0)}-\Delta_i S_{m+1}^{(0)}]   
+   
    (r'_{m-1}-r_m) \Delta_i S_{m-1}^{(0)}
    +(l'_{m+1}-l_m) \Delta_i S_{m+1}^{(0)} , 
\end{eqnarray}
\begin{multicols}{2}
where we have used that 
   $s_m=1-r_m-l_m $. 
Expressing $r_m$ and $l_m$ in terms of driving forces and transport parameters 
(\ref{eq:rmlm}) 
and keeping only leading order terms in $a$, one obtains
\begin{eqnarray}
   \delta \Delta_i S  \label{eq:dDiS} 
&=& 
   \tau D \dx^{2} (\Delta_i S^{(0)}) - \tau v \dx (\Delta_i S^{(0)}) \non\\
&-&
   \tau (\dx v) \Delta_i S^{(0)} + {\cal O}(\tau^2,a) .
\end{eqnarray}
After division by $a\tau$, the right-hand side of (\ref{eq:dDiS}) contains 
the rate of irreversible entropy production 
$\Delta_i S^{(0)}/a\tau$ 
and its spatial derivatives,
because $\Delta_i S^{(0)}/a \tau $ has a finite macroscopic limit. 
Consequently, the difference in the entropy production 
   $\delta \Delta_i S/(a\tau) $, 
vanishes in the macroscopic limit, when $\tau\rightarrow 0$. 

Since the temporal change of the level-0 coarse-grained 
entropy is (by definition)
unaltered by changes of the prescription for coarse graining  
\begin{eqnarray}
   \Delta^{(2)} S_m^{(0)} 
&=&
   S_m^{(0)''}-S_m^{(0)}  \non\\
&=&
   S_m^{(0)''}-S_m^{(0)'}+S_m^{(0)'}-S_m^{(0)}  \non\\
&=&
   \Delta S_m^{(0)'}+\Delta S_m^{(0)} , 
\end{eqnarray}    
also the entropy flux, i.e., the difference of 
the change of 
the coarse-grained entropy
and the irreversible entropy production, takes the same macroscopic
limit, regardless of the procedure for coarse graining. 

Note that the calculations given in this Appendix can be generalized in a
straightforward manner to account for averaging only after every $n$-th time 
steps on any finite level $\kappa$ of the partitioning of the cells
\cite{refine}. 
All these approaches only differ by terms that can be expressed as a
product of $\tau$ and spatial derivatives of the macroscopic rate of
irreversible entropy production. 
For larger $n$ and $\kappa$, more and higher derivatives appear.


\end{multicols}


\begin{references}
\bibitem{refine} 
An extension in this spirit for the case of thermostatted systems with
constant temperature can be found in \cite{GD2}.
\bibitem{consistency}
In kinetic theories the fields are related by the square
of the momentum variable.  
Here we show that consistency 
with thermodynamics can be achieved 
without referring to this relation.
\bibitem{Boltzmann}
A closer inspection reveals that the multibaker model mimics
certain features of the Boltzmann equation approach in
relaxation-time approximation. 
There $v_{el}=eE\tau_{relax} /m$, where m is the mass of the particles, 
and $\tau_{relax}$ stands for the relaxation time.  
This shows that the finiteness of the diffusion coefficient 
and the conductivity in our approach corresponds to the finiteness of 
the relaxation time in the kinetic theory.
\end{references}
\end{document}